# Plasmons in the van der Waals charge-density-wave material 2H-TaSe₂


Chaoyu Song[1,2], Xiang Yuan[1,3], Ce Huang[1], Shenyang Huang[1,2], Qiaoxia Xing[1,2], Chong Wang[1,2], Cheng Zhang[1,4], Yuangang Xie[1,2], Yuchen Lei[1,2], Fanjie Wang[1,2], Lei Mu[1,2], Jiasheng Zhang[1,2], Faxian Xiu[1,4,5], Hugen Yan[1,2]*

1. State Key Laboratory of Surface Physics and Department of Physics, Fudan University, Shanghai 200433, China

2. Key Laboratory of Micro and Nano Photonic Structures (Ministry of Education), Fudan University, Shanghai 200433, China

3. State Key Laboratory of Precision Spectroscopy, East China Normal University, Shanghai 200062, China.

4. Institute for Nanoelectronic Devices and Quantum Computing, Fudan University, Shanghai 200433, China.

5. Shanghai Research Center for Quantum Sciences, Shanghai 201315, China

*email: hgyan@fudan.edu.cn




**Abstract:**

**Plasmons in two-dimensional (2D) materials beyond graphene have recently gained much attention. However, the experimental investigation is limited due to the lack of suitable materials. Here, we experimentally demonstrate localized plasmons in a correlated 2D charge-density-wave (CDW) material: 2H-TaSe$_2$. The plasmon resonance can cover a broad spectral range from the terahertz (40 µm) to the telecom (1.55 µm) region, which is further tunable by changing thickness and dielectric environments. The plasmon dispersion flattens at large wave vectors, resulted from the universal screening effect of interband transitions. More interestingly, anomalous temperature dependence of plasmon resonances associated with CDW excitations is observed. In the CDW phase, the plasmon peak close to the CDW excitation frequency becomes wider and asymmetric, mimicking two coupled oscillators. Our study not only reveals the universal role of the intrinsic screening on 2D plasmons, but also opens an avenue for tunable plasmons in 2D correlated materials.**

With strong tunability and extraordinary light field confinement, plasmons in 2D materials show great promise in reconfigurable photonics[1-11]. In the long wavelength limit, the plasmon frequency of 2D free electron gas is proportional to $\sqrt{q}$ [4, 12], with $q$ as the wave vector. Such dispersion may give us an illusion that the plasmon frequency can go as high as one wishes. However, recent theoretical studies suggest that the plasmon dispersion in real 2D materials flattens universally due to the intrinsic dielectric screening from interband transitions[13-15], which is inevitable for almost every crystal. The flattened dispersion renders slow plasmon group velocity and facilitates plasmon localization[15]. Though with importance, the experimental verification of such flattened



dispersion remains elusive up to date.

2H-TaSe$_2$ belongs to a transition metal dichalcogenide (TMDC), which attracts much attention due to the appearance of CDW orders[16]. It exhibits a normal-incommensurate CDW phase transition at about $T_{C1}$ = 122 K, followed by an incommensurate-commensurate CDW phase transition at $T_{C2}$ = 90 K[16, 17]. Previously, the bulk plasmon of metallic TMDCs was studied by electron energy-loss spectroscopy (EELS) and the anomalous negative plasmon dispersion was observed[18, 19]. The interactions between CDW and plasmons were proposed as the physics origin of the negative dispersion[19-21]. In contrast, some later theoretical calculations and doping experiments revealed that the narrow d bands near the Fermi level may account for the negative dispersion[22-24]. All these studies based on EELS are for bulk plasmons, while the coupling between 2D plasmons and CDW excitations in TaSe$_2$ thin films is still unclear.

Here, we report the experimental studies of localized 2D plasmons in 2H-TaSe$_2$ van der Waals (vdW) thin films. We patterned the vdW thin films of 2H-TaSe$_2$ into ribbon arrays and measured the plasmon resonances by Fourier-transform infrared spectroscopy (FTIR). The plasmonic excitations correspond to the collective oscillations of carriers with the moving direction perpendicular to the ribbon[1], analogous to localized plasmons in traditional metallic nanostructures. Of course, propagating surface plasmons can be expected to show up in TaSe$_2$ thin films through momentum compensation techniques, such as an atomic force microscopy (AFM) tip in near field imaging experiments[2, 3]. We find that the resonance frequency of the plasmon in TaSe$_2$ thin films covers a broad spectra range and extends to the telecom region, which is



unattainable for graphene plasmons[6]. In addition, TaSe$_2$ plasmons sensitively depend on the layer thickness and dielectric environments. Particularly, we reveal that the plasmon dispersion becomes flat at large wave vectors due to the screening of interband transitions, fully consistent with theoretical predictions[15]. More interestingly, we find that the CDW phase has profound influence on the plasmon resonance. In our study, we observe the coupling effects between TaSe$_2$ plasmons and CDW excitations, which causes non-monotonic change of the peak height and linewidth when the temperature decreases. For comparison, thin films of 2H-NbSe$_2$ were fabricated into ribbon arrays as well. On the contrary to TaSe$_2$, the plasmon peak of NbSe$_2$ continually becomes sharper with decreasing temperature.

**Results**

**The 2D plasmon dispersion of 2H-TaSe$_2$**

2H-TaSe$_2$ exhibits hexagonal structures as illustrated in Fig. 1a. Thin films of TaSe$_2$ were mechanically exfoliated from bulk crystals onto diamond substrates, as displayed in Fig. 1b. In one uniform TaSe$_2$ film, we fabricated more than 10 plasmonic arrays with various ribbon widths. The extinction spectra, which are directly determined by the optical conductivity, were measured by FTIR as illustrated in Fig. 1c. When the polarization of incident light is along the ribbon direction, the extinction spectrum is dominated by free carrier response (Drude scattering rate 2000 cm$^{-1}$), as shown in Fig. 1d (sample T1, thickness $d$ = 40 nm, ribbon width $W$ = 500 nm), whereas plasmon resonances are observed with perpendicular polarization. The optical conductivity of plasmons is fitted by a Lorentz oscillator model (Eq. (4) - (5) in Methods).



Fig. 2a shows the extinction spectra of TaSe$_2$ plasmons for ribbon arrays with different widths. The first and the second order plasmon peaks are simultaneously observed for plasmonic devices with ribbon width larger than 500 nm. The resonance frequency of the first order plasmon increases significantly from 1035 cm$^{-1}$ to 6133 cm$^{-1}$ as the ribbon width decreases from 2 μm to 60 nm, which corresponds to the light wavelength $\lambda$ changes from 9.6 μm to 1.6 μm. Moreover, the plasmon resonance of TaSe$_2$ can cover terahertz and far-infrared (far-IR) regions as well, with the resonance light wavelength readily beyond 40 μm (see Supplementary Fig. 1). The relatively broad spectral range of TaSe$_2$ plasmons is originated from its high carrier density and the absence of Landau damping channels within this spectral range at room temperature. The effective sheet carrier density of the TaSe$_2$ film with thickness of 40 nm is estimated to be $3.8 \times 10^{17}$ cm$^{-2}$ at room temperature (see the Hall resistance measurement in Supplementary note 11), much larger than that of graphene ($2.5 \times 10^{13}$ cm$^{-2}$ for highly doped graphene[5]) and comparable to that of ultrathin gold films ($1.8 \times 10^{16}$ cm$^{-2}$ for $d$ = 3 nm[25]).



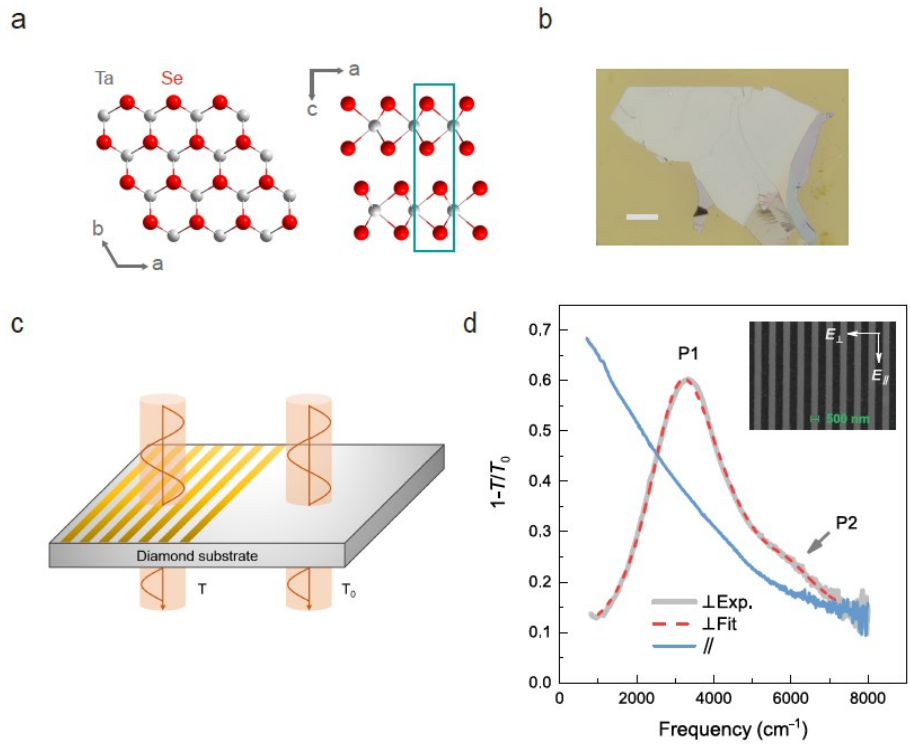

**Fig. 1 Characterization of 2H-TaSe$_2$ ribbon arrays. a** The crystal structures of 2H-TaSe$_2$ viewed along c-axis (monolayer) and b-axis directions (bilayer). The grey and red atoms represent tantalum (Ta) and selenium (Se) atoms respectively, and the cyan rectangle denotes the unit cell. **b** The optical image of a typical exfoliated TaSe$_2$ thin film with thickness of 40 nm. The scale bar is 100 μm. **c** The schematic of the setup for transmission measurements. The yellow stripes represent TaSe$_2$ ribbon arrays, $T_0$ and $T$ correspond to the transmission light of the bare diamond substrate and a ribbon array on the diamond substrate, respectively. **d** The extinction spectra 1-$T/T_0$ for a plasmonic device (sample T1, $d$ = 40 nm, $W$ = 500 nm), the blue and grey solid lines correspond to the light polarization parallel ( $/\!/$ ) and perpendicular ($\perp$) to the ribbon direction respectively. The red dashed line is the fitted curve of the plasmon peak. The first order (P1) and the second order (P2) plasmon peaks are denoted. The inset shows the scanning electron microscopy (SEM) image of the ribbon array.



In the long-wavelength limit, the plasmon dispersion of a 2D free electron system is given by[4, 12]:

$$\omega_p(q) = \sqrt{\frac{e^2}{2\varepsilon_0\varepsilon_e}\frac{n_s}{m}q} \quad (1)$$

where $\varepsilon_0$ is the vacuum permittivity, $\varepsilon_e$ is the relative dielectric constant of the surrounding environment, $n_s$ is the sheet carrier density, and $m$ is the effective mass of carriers. The wave vector $q$ is $\pi/W$ for the ribbon with width $W$[6]. As shown in Fig. 2b, the plasmon dispersion follows the $\sqrt{q}$ dependence at small wave vectors, whereas it becomes almost dispersionless when the wave vector is larger than $3\times10^7$ m$^{-1}$. As suggested by Jornada et al.[15], the deviation from the ideal 2D plasmon dispersion is originated from the screening effect of interband transitions, and the flattening of plasmon dispersion at large $q$ appears to be universal for 2D metals. For metallic TMDCs, there are multiple interband transitions between bands near the Fermi level, namely, the occupied (unoccupied) bands just below (above) the Femi level, whose onset energy is around 1-2 eV according to optical measurements[26-28] and first-principles calculations[14, 15, 22, 23]. The plasmon dispersion can be modified by introducing screening effects of interband transitions into the effective dielectric constant $\varepsilon_e$ [15], which becomes $q$-dependent in Keldysh model[29]. In the long-wavelength limit, the effective dielectric constant is approximately expressed as $\varepsilon_e = \frac{1+\varepsilon_s}{2} + \rho_0 q$, where $\varepsilon_s$ is the dielectric constant of the substrate, $\rho_0$ is the screening length of the 2D film. For thin films with finite thickness $d$, the screening length is $\rho_0 \approx d\varepsilon/2$, and $\varepsilon$ is the intrinsic dielectric constant due to interband transitions[29]. By incorporating the $q$-dependent dielectric constant $\varepsilon_e(q)$ into Eq. (1), the plasmon dispersion of 2D materials with interband screening reads[15]:

$$\omega_p(q) = \sqrt{\frac{n_s e^2}{2\varepsilon_0 m}\frac{q}{(1+\varepsilon_s)/2 + \rho_0 q}} \quad (2)$$



The plasmon dispersion as well as the saturation behavior at large wave vectors are well fitted by Eq. (2), as displayed in Fig. 2b. The fitted screening length $\rho_0$ is (710 ± 260) Å and the intrinsic dielectric constant $\varepsilon$ is about 3.5 ± 1.3 for 2H-TaSe$_2$ with thickness of 40 nm. The plasmon dispersion can also be simulated by the loss function[11] $-\mathrm{Im}(\dfrac{1}{\varepsilon(q,\omega)})$, as illustrated by the pseudo-color map in Fig. 2b, and more details can be found in Supplementary note 2. Plasmons at the flat dispersion region can exhibit low group velocity, strong light confinement and considerably large field enhancement[15], which holds promise for highly sensitive detection of local and non-local electronic structures.

As shown in Fig. 2c, the peak width of the first order plasmon is 552 cm$^{-1}$ at $1.6\times10^6$ m$^{-1}$ and increases to 3200 cm$^{-1}$ at $5.2\times10^7$ m$^{-1}$. Due to the retardation effect[30, 31], the plasmon peak width at small wave vectors is less than its Drude scattering rate. At relatively large wave vectors, where the retardation effect is already negligible, the plasmon width gradually increases with the wave vector increases. We attribute such peak width enhancement to the more pronounced edge scattering of carriers[6] and the inevitable nanofabrication imperfections and inhomogeneities.

In addition to the dispersion, the spectral weight of TaSe$_2$ plasmons is also strongly influenced by the interband screening. For the relatively wide ribbons, the spectral weight of the first order plasmon is less than that of the second order plasmon (see Fig. 2d), which is presumably transferred to the second order plasmon due to the retardation effect[30, 31]. Therefore, we focus on the relatively narrow ribbons whose the second order plasmon peak is nearly diminishing. That is the region with wave vector larger than $1\times10^7$ m$^{-1}$ where the plasmon spectral weight depends on



the wave vector as follows: $S_{\mathrm{P}} \propto (1+\alpha q)^{-1}$, where $\alpha$ is a coefficient associated to the intrinsic screening of interband transitions. This relation can be derived and generalized from the analytical solution of plasmons in metal disks[32, 33] (see Supplementary note 2). As shown in Fig. 2c, the spectral weight closely follows such scaling rule for $q > 1 \times 10^7$ m$^{-1}$. In contrast, for the ideal 2D free electron gas the plasmon spectral weight is independent of $q$. Thus, the decrease of plasmon spectral weights at large wave vectors is also a manifestation of the interband screening for real 2D materials.

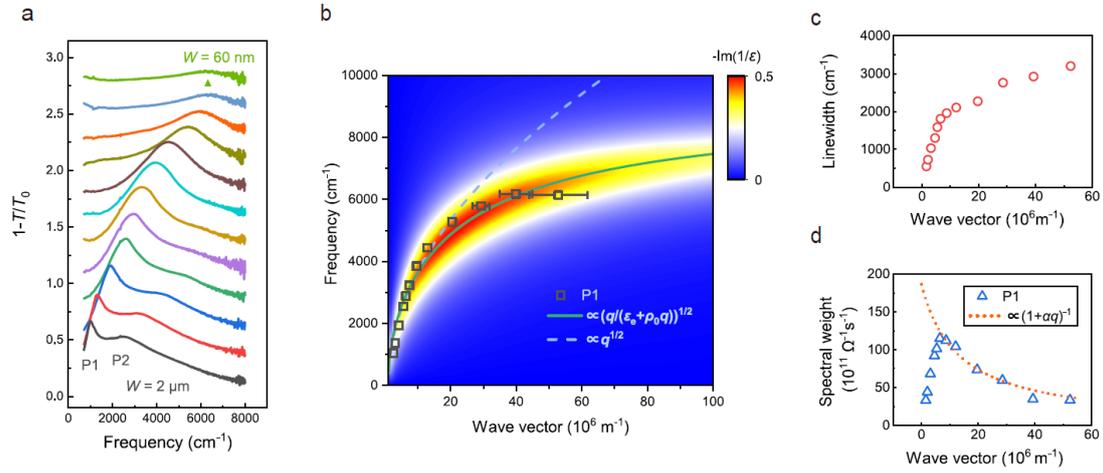

**Fig. 2 Plasmons in 2H-TaSe$_2$ thin films. a** The extinction spectra of plasmonic devices with ribbon width ranging from 60 nm to 2 μm, the spectra are vertically shifted for clarity. **b** The plasmon dispersion of a 2H-TaSe$_2$ thin film (sample T1, $d = 40$ nm). The black squares denote the frequencies of the first order plasmon, and the error bar represents the uncertainty from the inhomogeneity of ribbon width. The green solid line is the fitted dispersion based on Eq. (2). The blue dashed line is the $\sqrt{q}$ scaling to fit data points with small wave vectors (less than $1.5 \times 10^7$ m$^{-1}$). The pseudo-color map shows the calculated plasmon loss function $-\mathrm{Im}(1/\varepsilon)$. **c** The peak width and **d** the spectral weight of plasmon resonances. The orange dashed line is the fitted



spectral weight at large wave vectors which follows the $S_P \propto (1+\alpha q)^{-1}$ relation.

**Tunable plasmons by thickness and dielectric environments**

The plasmon resonance of 2D materials is tunable by changing thickness. For example, the intensity of graphene plasmons and the resonance frequency were effectively modified by stacking multiple layers of graphene[5]. To explore the thickness dependence of TaSe$_2$ plasmons, we fabricated plasmonic devices in TaSe$_2$ films with different thickness ranging from 10 nm to 40 nm. The intensity of the plasmon peak can be largely tuned by the thickness, as illustrated in Fig. 3a. In addition, when it is far away from the saturation frequency, the plasmon frequency is proportional to $\sqrt{d}$ according to Eq. (1), provided that $n_s \propto d$ and other material properties remain the same. This relation fits the plasmon frequency well, as shown in Fig. 3b. More details about the thickness dependence of TaSe$_2$ plasmons are shown in Supplementary Fig. 3.

TaSe$_2$ plasmons are also extremely sensitive to the surrounding dielectric environment. We fabricated TaSe$_2$ plasmonic devices on BaF$_2$ ($\varepsilon_s$ = 2.1, sample T5) and Si substrates ($\varepsilon_s$ = 11.8, sample T6), in addition to diamond substrates ($\varepsilon_s$ = 5.7). Low dielectric constant substrates facilitate the observation of higher energy plasmon in experiments. The plasmon frequency of the sample on BaF$_2$ substrates is about twice as high as that on Si, as illustrated in Fig. 3c. It perfectly follows Eq. (2), as indicated by the dashed curve in Fig. 3d. The fitted screening length $\rho_0$ for the film with thickness of 40 nm is (700 $\pm$ 280) Å and thus the intrinsic dielectric constant $\varepsilon$ is 3.5 $\pm$ 1.4, which agrees with the value fitted from the plasmon dispersion. The maximal plasmon frequency we observed is 6580 cm$^{-1}$ ($\lambda$ = 1.52 μm) for a plasmonic device on BaF$_2$ substrates



(sample T5, $d = 40$ nm, $W = 130$ nm). Therefore, the plasmon in TaSe$_2$ films covers the terahertz to the telecom range, which promises broad applications in photonics and optoelectronics. Note that, according to Eq. (2), the achievable maximal plasmon frequency is still limited by the intrinsic screening. More details on the substrate effect are displayed in Supplementary Fig. 4.

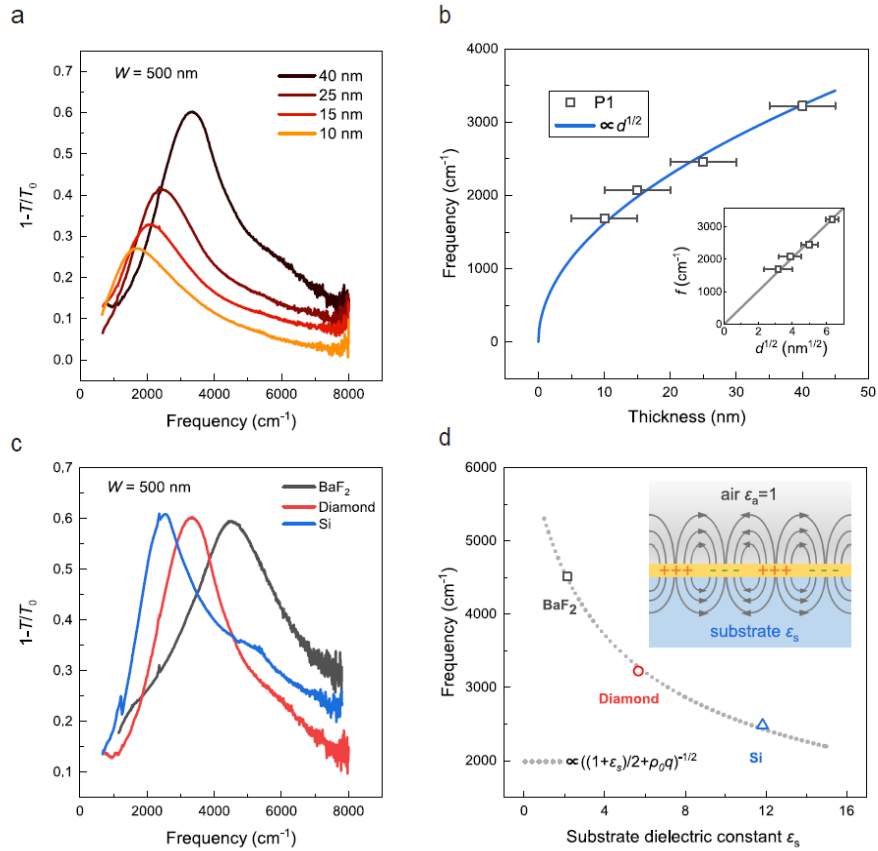

**Fig. 3 Tunable plasmons by thickness and dielectric environments. a** The extinction spectra of plasmonic devices with film thickness of 10 nm (an isolated sample), 15 nm (sample T3), 25 nm (sample T2) and 40 nm (sample T1) respectively. The ribbon width is fixed at 500nm. **b** The thickness dependence of the plasmon frequency, the blue solid line shows the fitted plasmon frequency as a function of $\sqrt{d}$, the grey solid line in the inset also shows this relation. **c** The extinction spectra of plasmonic devices on BaF$_2$ (sample T5), diamond (sample T1) and Si (sample T6) substrates respectively, with the same thickness and ribbon width ($d = 40$ nm, $W =$



500 nm). **d** The plasmon frequency as a function of substrate dielectric constants $\varepsilon_s$, the grey dashed line represents the fitted plasmon frequency which is proportional to $((1 + \varepsilon_s) / 2 + \rho_0 q)^{-1/2}$. The inset illustrates the screening effect of the substrate. The TaSe$_2$ thin film (yellow film) is surrounded by the substrate (blue region) and air (grey region) with dielectric constants of $\varepsilon_s$ and $\varepsilon_a$ respectively. The grey solid lines denote the electric field lines caused by the spatial distribution of positive charges (+) and negative charges (-) in the film.

**Plasmons and the CDW phase transition**

In general, the carrier scattering in metals is suppressed as the temperature decreases. As a consequence, the linewidth of the plasmon resonance reduces. For example, the WTe$_2$ plasmon peak becomes much sharper at cryogenic temperatures[11]. However, the plasmon of 2H-TaSe$_2$ exhibits anomalous temperature dependence. As shown in Fig. 4a and 4b, the TaSe$_2$ plasmon peak (sample T4, $W$ = 700 nm, $d$ = 20 nm) firstly becomes sharper and the peak height increases when the temperature decreases from 300 K to 120 K. Then, on the contrary, as the temperature drops below the CDW phase transition temperature $T_{C1}$, the plasmon peak becomes wider and asymmetric, and the peak height decreases. This is in sharp contrast to 2H-NbSe$_2$, the plasmon peak of which (sample N1, $W$ = 700 nm, $d$ = 25 nm) continually becomes sharper as the temperature decreases, as shown in Fig. 4c. The plasmon properties of 2H-NbSe$_2$ are generally similar to those of 2H-TaSe$_2$ except for the temperature dependence. The detailed study for NbSe$_2$ plasmons is presented in Supplementary note 6.

To understand the underlying mechanism of the abnormal temperature dependence of TaSe$_2$



plasmons, we measured its intrinsic optical response with incident light polarized along the ribbon direction. As shown in Fig. 4d, an absorption peak emerges as the temperature decreases below $T_{C1}$, while only Drude response is observed at higher temperature. The frequency of the absorption peak is 2180 cm$^{-1}$ (270 meV) and the peak width is approximately 1000 cm$^{-1}$ (124 meV) at 15K, as determined from a Lorentz fitting (see Supplementary Fig. 5). Meanwhile, the spectral weight of Drude response at this spectral range is suppressed and shifts to lower frequency range[26], which is a typical consequence of opening a partial gap[34]. Based on the above analysis, we attribute the excitation peak to the opening of a partial CDW gap, whose size is comparable to that of bulk TaSe$_2$, which is 150-250 meV determined by different techniques such as infrared reflection spectroscopy[26, 27, 35], angle-resolved photoemission spectroscopy (ARPES)[36-39] and scanning tunneling microscopy (STM)[40]. Furthermore, our electrical transport measurements on a TaSe$_2$ thin film also corroborate the appearance of CDW phase transition: the slope of the resistivity changes at $T_{C1}$ = 122 K as illustrated in the inset of Fig. 4d. Note that the CDW order preserves for monolayer and few-layer metallic TMDCs[41, 42], and for flakes thicker than 10 nm the phase transition temperature remains almost the same as that of the bulk [43], which is the case in our study. For 2H-NbSe$_2$, no clear CDW excitation is observed in previous reflection measurements[28], presumably due to the competition between the superconductivity ($T_S$ = 7.2 K) and CDW orders ($T_C$ = 33 K)[16, 17]. This is consistent with our measurements on the intrinsic optical response of 2H-NbSe$_2$, as detailed in Supplementary note 7.



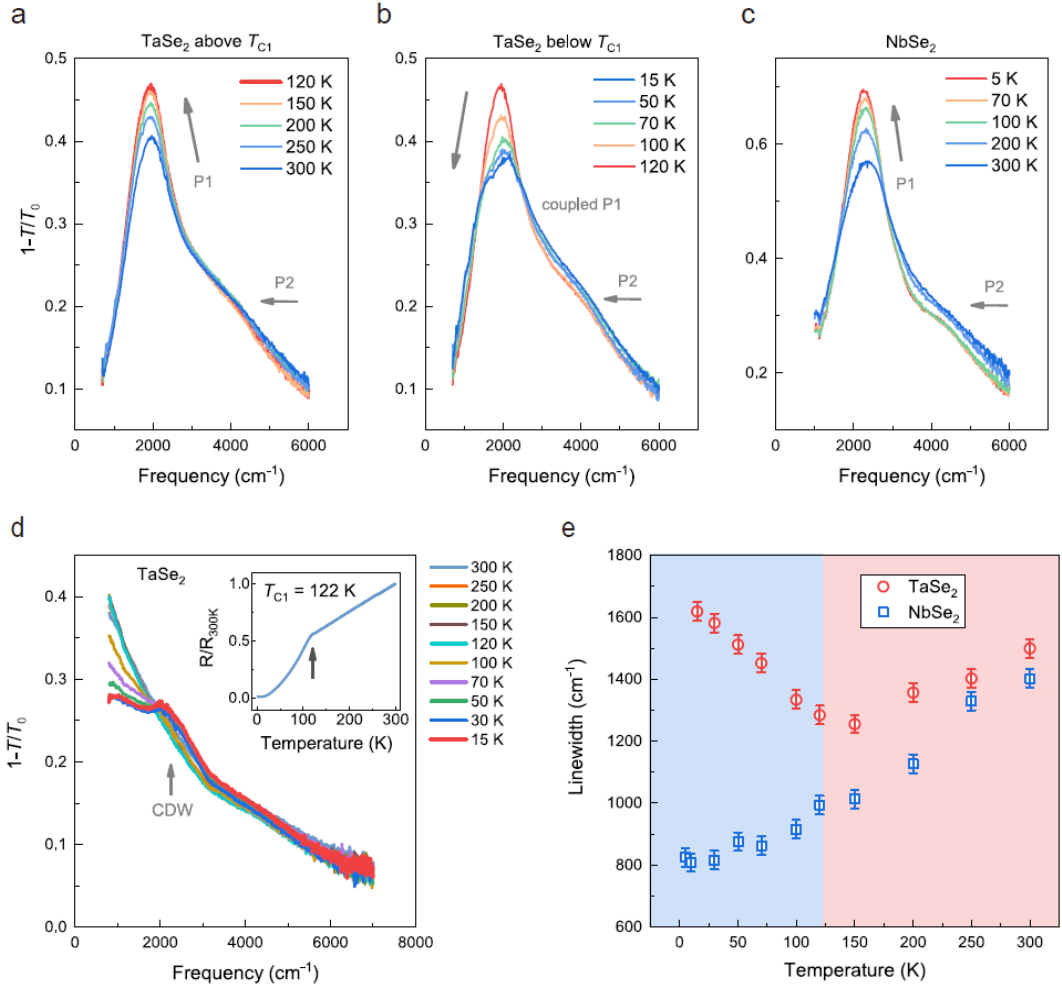

**Fig. 4 The temperature dependence of 2H-TaSe$_2$ and 2H-NbSe$_2$ plasmons. a** The temperature evolution of the plasmon spectra of 2H-TaSe$_2$ (sample T4, $W$ = 700 nm, $d$ = 20 nm) above $T_{C1}$. **b** same as **a** but below $T_{C1}$. **c** The temperature-dependent plasmon spectra of 2H-NbSe$_2$ (sample N1, $W$ = 700 nm, $d$ = 25 nm). The arrows in **a-c** beside the main peaks indicate the peak height evolution with decreasing temperature, P1 and P2 denote the first and the second order plasmons respectively. **d** The extinction spectra of 2H-TaSe$_2$ with light polarized along the ribbon direction at various temperatures. The CDW excitation appears with temperature below $T_{C1}$. The inset shows the normalized temperature-dependent resistivity of a TaSe$_2$ thin film ( $d \approx$ 20 nm). **e** The fitted temperature-dependent peak widths of TaSe$_2$ and NbSe$_2$ plasmons (the first order), the error bars show the uncertainty of spectra fitting. The pink and blue backgrounds denote the normal



metal and CDW phases of 2H-TaSe$_2$, respectively.

After identification of the CDW excitation, now we can attribute the anomalous plasmon behavior to the coupling between plasmonic and CDW excitations, since the plasmon frequency is in the vicinity of the CDW resonance frequency. A phenomenological coupled-oscillator model is applied to describe the interplay between them[44] (more details are presented in Supplementary note 8). The optical conductivity of the plasmon resonance coupled with a CDW excitation can be expressed as follows:

$$\sigma_1(\omega) = -i\frac{D_1}{\pi}\frac{\omega(\omega_2^2 - \omega^2 - i\gamma_2\omega)}{(\omega_1^2 - \omega^2 - i\gamma_1\omega)(\omega_2^2 - \omega^2 - i\gamma_2\omega) - \Omega^4} \quad (3)$$

Here $\omega_1$, $\omega_2$, $\gamma_1$, $\gamma_2$ correspond to the frequencies and the damping rates of oscillator 1 and 2 respectively, and $\Omega$ is the coupling rate between the two oscillators. Oscillator 1 represents the TaSe$_2$ plasmonic excitation, which can be directly excited by the driving electric field $E(t) = E_0 e^{-i\omega t}$, while oscillator 2 represents the CDW excitation. The CDW excitation is much weaker than the plasmonic excitation (see Supplementary Fig. 5), hence we treat it as a quasi "dark" mode in the model, which can be excited only indirectly through the coupling with the bright plasmon mode[45].

The first order plasmon of TaSe$_2$ below $T_{C1}$ is fitted by Eq. (3). While for simplicity, the second order plasmon is still fitted by the uncoupled oscillator model as Eq. (5), because it is relatively weak and far away from the frequency of the CDW excitation. Fig. 4e shows the temperature evolution of the peak width of TaSe$_2$ and NbSe$_2$ plasmons ($W$ = 700 nm). Above $T_{C1}$, the peak width of the first order TaSe$_2$ plasmons decreases from 1500 cm$^{-1}$ (300 K) to 1280 cm$^{-1}$ (120 K).



However, when the temperature drops below $T_{C1}$, the linewidth of the plasmon peak increases as the temperature decreases and eventually reaches 1620 cm$^{-1}$ at 15 K, in sharp contrast to the behavior of NbSe$_2$ plasmons. The temperature dependence of the peak width of 2H-TaSe$_2$ plasmons deviates from that of its Drude scattering rate, which continually reduces as the temperature decreases[26].

To reveal the coupling effect further, we fabricated multiple plasmonic devices with plasmon resonance frequency sweeping across the CDW excitation (sample T4, $W$ = 900-400 nm, $d$ = 20 nm). Fig. 5a shows the extinction spectra of them at 120 K and 15 K. The detailed temperature evolution of the plasmon peak is displayed in Supplementary Fig. 5. For extinction spectra at 15 K, as ribbon width decreases, the shape of the resonance peak continuously evolves due to the coupling to the CDW excitation. More prominent coupling effect is observed when the frequency of the plasmon and the CDW excitation is closer, such as the plasmonic devices with ribbon width of 800-600 nm. When the plasmon frequency is away from the CDW excitation, the broadening and asymmetry of plasmon peaks are less pronounced. The coupling effect becomes negligible at the far-IR region (see Supplementary Fig. 1), where the plasmon peak continually sharpens as the temperature drops, similar to that of 2H-NbSe$_2$ in uncoupled situations.



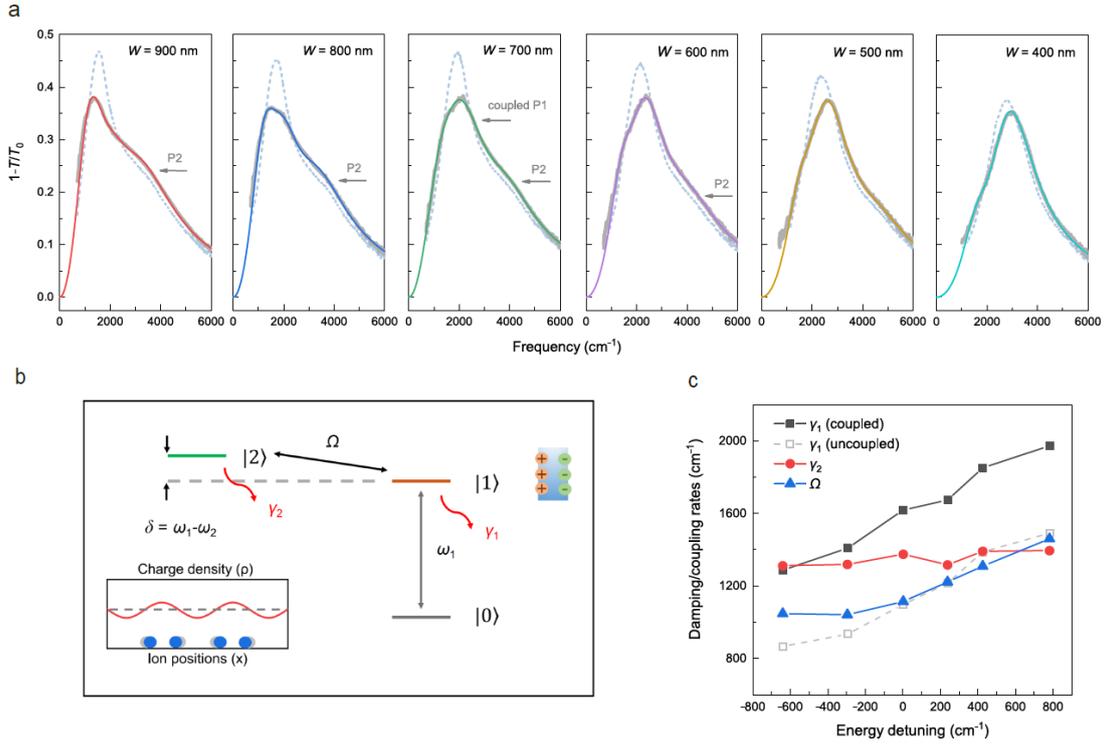

**Fig. 5 The coupling between plasmonic and CDW excitations in 2H-TaSe$_2$. a** The extinction spectra of TaSe$_2$ plasmons with ribbon width of 900-400 nm. The blue dashed and the grey solid lines correspond to the measured spectra at 120 K and 15 K, respectively. The colored lines represent the fitted spectra at 15 K by the coupled-oscillator model. P1 and P2 denote the first and the second order plasmons respectively. **b** A sketch for the coupling effect in a prototype three-level system. The grey, orange and green solid lines illustrate the energy level of the ground state $|0\rangle$, the bright excited state $|1\rangle$ and the dark state $|2\rangle$ respectively. The grey arrow denotes transition 1 $(|0\rangle \rightarrow |1\rangle)$ with frequency of $\omega_1$. The grey dashed line is plotted to define the energy detuning $\delta = \omega_1 - \omega_2$, where $\omega_2$ is the transition frequency between state $|2\rangle$ and $|0\rangle$. The red arrows near the state $|1\rangle$ and $|2\rangle$ represent the damping rates of the corresponding transitions, which are $\gamma_1$ and $\gamma_2$ respectively. The black arrow defines the coupling rate $\Omega$ between state $|1\rangle$ and $|2\rangle$. The left inset illustrates the modulation of charge density ($\rho$) and ion positions ($x$) in the CDW phase. The right inset denotes the positive (orange



circle) and negative charge (green circle) in the ribbon when the localized plasmon is excited. **c** The fitted damping/coupling rates as a function of energy detuning (the unit is wave number) at 15 K. The uncoupled plasmon damping rate $\gamma_1$ is estimated by the extrapolation from the temperature evolution of the plasmon peak width above $T_{C1}$, which generally follows linear relations (see Supplementary Fig. 5).

## Discussion

Depending on the coupling strength, the coupling effects between plasmons and other excitations can behave in different forms such as Purcell effect[46], Fano resonance[47, 48], plasmonic analogue of electromagnetically induced transparency (EIT)[45, 49-53] and Rabi splitting[54-58]. The coupling between two subsystems can enhance light-matter interactions and play an important role in applications such as nanolasers[59], sensors[60] and quantum information processing[61]. The coupling effect between TaSe$_2$ plasmonic and CDW excitations is originated from the interference effect between two different excitation pathways. Fig. 5b shows the scheme for the coupling effect in a prototype three-level system. The transition between the energy level $|1\rangle$ and $|0\rangle$ with resonance frequency $\omega_1$ can be directly excited, which corresponds to the plasmon excitation. Whereas the transition between energy level $|2\rangle$ and $|0\rangle$ is forbidden or extremely weak, which represents the quasi "dark" CDW excitation[45]. The energy detuning between the two transitions is $\delta = \omega_1 - \omega_2$. There are two different excitation pathways contributing to the occupation of the energy level $|1\rangle$, i.e., $|0\rangle \rightarrow |1\rangle$ and $|0\rangle \rightarrow |1\rangle \rightarrow |2\rangle \rightarrow |1\rangle$. The two excitation pathways interfere with each other, leading to a modification of the extinction spectrum. The fitted frequencies of the plasmonic ($\omega_1$) and the CDW excitations ($\omega_2$) as a function of



wave vectors (sample T4, $W$ = 900-400 nm, $d$ = 20 nm, $T$ = 15 K) are depicted in Supplementary

Figure 8b. Generally, their dispersions are approximately close to those in uncoupled situations.

Fig. 5c presents the damping/coupling rates fitted by the coupled-oscillator model as a function of

energy detuning. The plasmon damping rate $\gamma_1$ is enhanced compared to the uncoupled

conditions. Meanwhile, the damping rate of the CDW excitation $\gamma_2$ is also enhanced from 1000

cm$^{-1}$ in the uncoupled case to 1300-1400 cm$^{-1}$. The coupling rate $\Omega$ increases as the energy

detuning increases from negative to positive. It presumably suggests an increase of the effective

density of CDW excitations[62] coupled to the plasmonic field as the size of nanostructures

decreases.

The minimal energy detuning is realized for the plasmonic device with ribbon width of 700 nm. In

that case, $\gamma_1$, $\gamma_2$ and $\Omega$ are 1618 cm$^{-1}$, 1375 cm$^{-1}$ and 1114 cm$^{-1}$ respectively. With these

parameters in hand, we can make a comparison between the coupling effect in 2H-TaSe$_2$ and two

typical coupling phenomena in plasmonic systems. The first is the classical analogue of EIT[45, 49-51].

In plasmonic EIT systems, the damping rate of the dark mode $\gamma_2$ is significantly smaller than

that of the bright mode $\gamma_1$, such as the non-radiative quadrupole mode versus radiative dipole

mode in metamaterials[45, 49, 50] and the plasmon-phonon coupled systems[51, 52]. In our case, $\gamma_2$ of

the quasi "dark" CDW excitation is comparatively large. Consequently, only modification of the

lineshape instead of a sharp dip is observed in extinction spectra. However, if we can effectively

reduce $\gamma_2$, a dip can emerge, as simulated in Supplementary Fig. 8. It implies that the coupling

effects in TaSe$_2$ and EIT plasmonic systems share the same physics origin. The second is the

strong Rabi splitting in exciton-plasmon systems[55-58]. In those systems, the coupling rate $\Omega$ is



larger than the damping rates of the two excitations, which results in the formation of half-matter, half-light exciton-plasmon polaritons. Meanwhile, the hybridized polariton dispersion typically exhibits pronounced anti-crossing behavior, with large splitting in the dispersion intersection area. Here, the coupling rate $\Omega$ between TaSe$_2$ plasmonic and CDW excitations is less than their damping rates and no clear anti-crossing scenario can be observed.

In light of the coupling effects, plasmons can be utilized in probing and manipulating other polaritons like excitons[63, 64], phonons[51, 52], excitations in correlated materials[65] and non-local quantum response of neighbouring metals[66]. However, the coupling effect in 2H-TaSe$_2$ have something unique. Plasmonic and CDW excitations are both internal excitations, while in many other coupled plasmonic systems, the coupling depends on excitations in different materials or structures, such as excitons in semiconductor layers and plasmons in metal nanostructures[55-58]. Therefore, TaSe$_2$ plasmons can in return effectively manifest its intrinsic excitations. Once the plasmon peak becomes damped or asymmetric, it indicates there might exist intrinsic excitations at nearby frequency. This can open an avenue in probing weak excitations in other phase transition or correlated systems[67].

In summary, we experimentally demonstrated plasmons in 2H-TaSe$_2$ thin films, which covers a broad spectral range from the terahertz to the near-infrared region. The plasmon dispersion flattens at large wave vectors due to the screening of interband transitions. TaSe$_2$ plasmons are tunable by varying the film thickness and dielectric environments. In addition, the coupling effect between TaSe$_2$ plasmonic and CDW excitations is observed. The interplay between them is well interpreted



by the coupled-oscillator model. 2H-TaSe$_2$, along with other metallic TMDCs, has been demonstrated as a competitive building block in photonics. Our study can stimulate further endeavors on the plasmon in atomically thin TMDCs, given that the CDW phase transition temperature of monolayer and few-layer metallic TMDCs are dramatically different than that of their relatively thick or bulk counterparts[41]. Moreover, large-scale atomically thin TMDCs are available from controllable synthesis by chemical vapor deposition (CVD)[68, 69] or molecular beam epitaxy (MBE)[42] methods, which can render more tunabilities such as electrical gating[1] or chemical doping[5].

## Methods

### Sample preparation and nanofabrication

The bulk single crystals of 2H-TaSe$_2$ were grown by a chemical vapor transport (CVT) method[18]. The 2H-NbSe$_2$ single crystals were bought from 2D Semiconductors Inc. We checked the crystal quality by Raman spectroscopy[41, 70] (see Supplementary Fig. 9). The thin films were mechanically exfoliated onto polydimethylsiloxane (PDMS) substrates from bulk crystals and then transferred to diamond or other substrates[11]. Diamond substrates are used in this work unless otherwise stated. The film thickness was determined by a stylus profiler (Bruker DektakXT). The lateral size of one thin film with uniform thickness is at most 200-300 μm. For mid-infrared (mid-IR) measurements, we can fabricate 6-12 plasmonic arrays with different ribbon widths in one homogeneous thin film. As such, the thickness uncertainty can be excluded in the determination of the plasmon dispersion. For far-IR measurements, one thin film is only large enough for one plasmonic array, because the beam size of the focused far-IR light is larger. We used electron beam lithography and reactive ion



etching (the reaction gas is $CF_4$) to fabricate ribbon arrays. Scanning electron microscope (Zeiss Sigma SEM) was used to finally determine the ribbon width. In general, the lateral etching is on the same order of magnitude of film thickness. No notable oxidation and degradation were found during the process of device fabrications and optical measurements, for the relatively thick TMDC films studied in this work (see Supplementary Fig. 10).

**Optical measurements and the fitting of extinction spectra**

We used a Bruker FTIR spectrometer (Vertex 70v) in conjunction with a Hyperion 2000 microscope to measure the extinction spectra. The incident light was focused on samples with a $15\times$ IR objective. A liquid-nitrogen-cooled Mercury-Cadmium-Telluride (MCT) detector and a ZnSe grid polarizer were used for mid-IR measurements, while a liquid-helium-cooled silicon bolometer and a terahertz polarizer were applied in far-IR measurements. The low-temperature measurements were carried out with a helium-flow cryostat (Janis Research ST-300).

The relation between the extinction spectrum and the sheet optical conductivity $\sigma(\omega)$ is given by[5]:

$$1 - \frac{T}{T_0} = 1 - \frac{1}{\left| 1 + Z_0 \sigma(\omega) / (1 + n_{sub}) \right|^2} \qquad (4)$$

Where $Z_0$ is the vacuum impedance, $n_{sub}$ is the refractive index of the substrate, which is 2.38 for diamond substrates in the mid-IR range (1.46 for $BaF_2$ and 3.44 for Si substrates), and $\omega$ is the light frequency. The optical conductivity contributed from the plasmon (without coupling to other excitations, such as CDW) is[5]:



$$\sigma(\omega) = i \frac{f \cdot S_P}{\pi} \frac{\omega}{\omega^2 - \omega_p{}^2 + i\gamma_p\omega} \qquad (5)$$

Where $S_P$, $\omega_P$ and $\gamma_P$ are the spectral weight, the frequency and the damping rate of plasmon resonance respectively, and $f$ is the filling factor (the area of nanoribbons/all areas). The designed filling factor of our plasmonic devices is 0.4-0.5, but the lateral etching in fabrication process usually reduces the actual filling factor. Therefore, it is determined by the real ribbon width measured by SEM.

**Electrical transport measurements**

The TaSe$_2$ and NbSe$_2$ flakes were mechanically exfoliated onto Si/SiO$_2$ substrates. The electrical devices were fabricated by electron-beam lithography technique and wet-etched by standard buffered HF solution for 5 s in the electrode regime. We deposited 5 nm-thick Cr and 80 nm-thick Au electrodes using magnetic sputtering. Four-terminal temperature-dependent transport measurements were carried out in a Physical Property Measurement System (PPMS, Quantum Design) which achieved a base temperature of 1.8 K, in conjunction with lock-in amplifiers (SR830).

**Data availability**

The data that support the findings of this study are available from the corresponding author upon reasonable request.

**Acknowledgements**

H. Y. is grateful to the financial support from the National Key Research and Development Program of China (Grant Nos. 2016YFA0203900 and 2017YFA0303504), the National Natural Science Foundation of China (Grant Nos. 12074085 and 11734007), the Natural Science Foundation of Shanghai (Grant No. 20JC1414601) and Strategic Priority Research Program of Chinese Academy of Sciences (No. XDB30000000). F. X. is grateful to the financial support from the National Natural Science Foundation of China (Grant Nos. 11934005, 61322407 and 11874116), the Science and Technology Commission of Shanghai (Grant No. 19511120500) and the Program of Shanghai Academic/Technology Research Leader (Grant No. 20XD1400200). X. Y. acknowledges the financial support from the Science and Technology Commission of Shanghai (Grant Nos. 20YF1411700 and 20520710900). C. W. is grateful to the financial support from the National Natural Science Foundation of China (Grant No. 11704075). The authors thank L. J. Wang, W. Li, Z. Z. Sun and J. M. Luo for the help in experiments. Part of the experimental work was carried out in Fudan Nanofabrication Lab.

**Author contributions**

H. Y. conceived the research. C. S. performed sample preparation, optical measurements and data analysis with assistance from S. H., Q. X., C. W., Y. X., Y. L., F. W., L. M. and J. Z.; F. X., X. Y. and C. Z. grew and characterized the bulk single crystals of 2H-TaSe$_2$; C. H. performed electrical measurements; H. Y. and C. S. cowrote the paper; all authors commented on the paper.

**Competing interests**



The authors declare no competing interests.

**Additional information**

**Supplementary information** is available for this paper at https://doi.org/

**Correspondence** and requests for materials should be addressed to H. Y.

**Peer review information** Nature Communications thanks reviewer(s) for their contribution to the peer review of this work.

**Reprints and permission information** is available at http://www.nature.com/reprints

**Publisher's note** Springer Nature remains neutral with regard to jurisdictional claims in published maps and institutional affiliations